\documentclass{article}

\usepackage[hidelinks]{hyperref}

\usepackage[margin=1in]{geometry}

\usepackage{setspace}

\usepackage{booktabs, tabularx, adjustbox, multirow}

\usepackage{blkarray}

\usepackage{graphicx}
\usepackage[nomarkers, nolists]{endfloat}

\usepackage[round, authoryear]{natbib}
\usepackage{amsmath,amssymb}
\usepackage{mathrsfs}

\newcommand{\tS}{{\textrm{S}}}
\newcommand{\tI}{{\textrm{I}}}
\newcommand{\tB}{{\textrm{B}}}
\newcommand{\tR}{{\textrm{R}}}

\newcommand{\bT}{{\boldsymbol{T}}}
\newcommand{\bt}{{\boldsymbol{t}}}

\newcommand{\btheta}{{\boldsymbol{\theta}}}
\newcommand{\bpi}{{\boldsymbol{\pi}}}

\newcommand{\mC}{{\mathscr{C}}}

\newcommand{\mT}{{\mathscr{T}}}

\newcommand{\mcM}{{\mathcal{M}}}
\newcommand{\mcS}{{\mathcal{S}}}

\newcommand{\simsamplesizefig}{Fig.\ S1}
\newcommand{\posteriorlearning}{Fig.\ S2}
\newcommand{\modelfit}{Fig.\ S3}
\newcommand{\simconfigurations}{Table S3}
\newcommand{\countytables}{Tables S4--6}

\title{A method for characterizing disease emergence curves from paired pathogen detection and serology data}

\usepackage{authblk}

\author[1*]{Joshua Hewitt}
\author[1]{Grete Wilson-Henjum}
\author[3]{Derek T. Collins}
\author[3]{Jourdan M. Ringenberg}
\author[5]{Christopher A. Quintanal}
\author[5]{Robert Pleszewski}
\author[5]{Jeffrey C. Chandler}
\author[4]{Thomas J. DeLiberto}
\author[2]{Kim M. Pepin}

\affil[1]{Department of Wildland Resources, Utah State University, Logan, UT, USA}
\affil[2]{National Wildlife Research Center, Wildlife Services, Animal and Plant Health Inspection Service, United States Department of Agriculture, Fort Collins, CO, USA}
\affil[3]{National Wildlife Disease Program, Wildlife Services, Animal and Plant Health Inspection Service, United States Department of Agriculture, Fort Collins, CO, USA}
\affil[4]{Wildlife Services, Animal and Plant Health Inspection Service, United States Department of Agriculture, Fort Collins, CO, USA}
\affil[5]{Wildlife Disease Diagnostic Laboratory, Wildlife Services, Animal and Plant Health Inspection Service, United States Department of Agriculture, Fort Collins, CO, USA}
\affil[*]{Corresponding author: Joshua Hewitt, 4101 Laporte Ave, Fort Collins, CO, 80521, josh.hewitt@usu.edu}
\date{}

\begin{document}

    
\author[1*]{Joshua Hewitt}
\author[1]{Grete Wilson-Henjum}
\author[3]{Derek T. Collins}
\author[3]{Jourdan M. Ringenberg}
\author[5]{Christopher A. Quintanal}
\author[5]{Robert Pleszewski}
\author[5]{Jeffrey C. Chandler}
\author[4]{Thomas J. DeLiberto}
\author[2]{Kim M. Pepin}

\affil[1]{Department of Wildland Resources, Utah State University, Logan, UT, USA}
\affil[2]{National Wildlife Research Center, Wildlife Services, Animal and Plant Health Inspection Service, United States Department of Agriculture, Fort Collins, CO, USA}
\affil[3]{National Wildlife Disease Program, Wildlife Services, Animal and Plant Health Inspection Service, United States Department of Agriculture, Fort Collins, CO, USA}
\affil[4]{Wildlife Services, Animal and Plant Health Inspection Service, United States Department of Agriculture, Fort Collins, CO, USA}
\affil[5]{Wildlife Disease Diagnostic Laboratory, Wildlife Services, Animal and Plant Health Inspection Service, United States Department of Agriculture, Fort Collins, CO, USA}
\affil[*]{Corresponding author: josh.hewitt@usu.edu}


\maketitle


\section*{Abstract}

\begin{enumerate}

    \item Wildlife disease surveillance programs and research studies track infection and identify risk factors for wild populations, humans, and agriculture.  Often, several types of samples are collected from individuals to provide more complete information about an animal's infection history.  Methods that jointly analyze multiple data streams to study disease emergence and drivers of infection via epidemiological process models remain underdeveloped.  Joint-analysis methods can more thoroughly analyze all available data, more precisely quantifying epidemic processes, outbreak status, and risks.
    
    \item We contribute a paired data modeling approach that analyzes multiple samples from individuals.  We use ``characterization maps'' to link paired data to epidemiological processes through a hierarchical statistical observation model.  Our approach can provide both Bayesian and frequentist estimates of epidemiological parameters and state.  Our approach can also incorporate test sensitivity and specificity, and we propose model fit diagnostics.   We motivate our approach through the need to use paired pathogen and antibody detection tests to estimate parameters and infection trajectories for the widely applicable susceptible, infectious, recovered (SIR) model. We contribute general formulas to link characterization maps to arbitrary process models and datasets and an extended SIR model that better accommodates paired data. 
     
    \item We find via simulation that paired data can more efficiently estimate SIR parameters than unpaired data, requiring samples from 5--10 times fewer individuals.  We use our method to study SARS-CoV-2 in wild White-tailed deer (\emph{Odocoileus virginianus}) from three counties in the United States.  Estimates for average infectious times corroborate captive animal studies.  The estimated average cumulative proportion of infected deer across the three counties is 73\%, and the basic reproductive number ($R_0$) is 1.88. 
    
    \item Wildlife disease surveillance programs and research studies can use our methods to jointly analyze paired data to estimate epidemiological process parameters and track outbreaks.  Paired data analyses can improve precision and accuracy when sampling is limited.  Our methods use general statistical theory to let applications extend beyond the SIR model we consider, and to more complicated examples of paired data.  The methods can also be embedded in larger hierarchical models to provide landscape-scale risk assessment and identify drivers of infection. 

\end{enumerate}

\noindent Keywords: data fusion, epidemiological process model, hierarchical modeling, paired data

\section{Introduction}

Understanding the process of disease emergence and drivers of infection hotspot dynamics are central aims in disease ecology \citep{manlove2022}.  Predicting how disease can spread across a landscape can help evaluate potential impacts to wild populations, identify risk for people and agriculture, and provide means to evaluate potential intervention strategies. Since individual outbreaks may move quickly and data may be challenging to collect, methods should focus on collecting multiple sample types from individuals.  Some wildlife disease surveillance programs and research studies use several types of samples from each individual to monitor an infectious disease in a population \citep{borremans2016,bevins2023,mcbride2023}.  However, methods that can analyze multiple sample types simultaneously to learn about a single infection process are relatively uncommon \citep[e.g.,][]{borremans2016,prager2020}.  We focus on data captured by assays for active virus detection (polymerase chain reaction; PCR) and antibody detection (serum) and contribute a statistical framework that simultaneously links both types of samples to a single infectious process.  This statistical framework is formally more general, as it can accommodate more than two sample types and be applied to many infectious process models.  The contributed methods can increase the precision with which surveillance programs can monitor an infectious disease when programs collect several types of data from each individual.

Wildlife disease surveillance programs and research studies commonly incorporate serological sampling (i.e., testing for antibodies) as a cost-effective way to answer questions about disease emergence mechanisms.  Serological sampling indirectly measures infection transmission levels, but provides a longer window of time to detect potentially rare transmission events since antibodies typically persist longer than the active phase of infection \citep{gilbert2013,wilber2020,bevins2021}.  Surveillance programs and methods that lean on opportunistic PCR and serology sampling for large-scale disease surveillance remain mainstream \citep[e.g.,][]{wilber2020,nichols2021,bevins2023,mysterud2023}.

Surveillance programs may also collect several biomarkers from each individual to more thoroughly evaluate infection and study emergence processes, including samples for pathogen detection and clinical signs of disease \citep[e.g.,][]{prager2020,bevins2023,mcbride2023}. The most direct infection transmission rate estimates incorporate time-of-infection data, which often requires multiple types of samples to quantify.   Methods that jointly analyze several types of samples to estimate time of infection tend to use antibody quantity data and prior knowledge about antibody dynamics.  Antibody quantities can be informative for infection rates via back-calculated time of infection estimates \citep{simonsen2009,teunis2012,borremans2016,pepin2017,prager2020}.  But, not all surveillance programs produce antibody quantity data, and prior knowledge about antibody dynamics may not be available for emerging infectious diseases in of species concern.  Methods that jointly analyze several types of samples without antibody quantities are underdeveloped.

Surveillance programs may analyze data with epidemiological process models to study emergence dynamics and trends.  Epidemic process models are mechanistic models that track which individuals are susceptible, infectious, or recovered from an infection, known as the SIR model \citep{anderson1991}.  The SIR and related models share the common concept of individuals transitioning between infection states, which enables inference of the population's epidemiological state over time. The infectious state is conventionally assessed by sampling a host for pathogen presence at a known route of pathogen shedding. Similarly, recovery from infection is assessed by assaying for specific antibodies to a pathogen in a serum sample. Hosts are otherwise assumed to be susceptible to infection.  It may not be possible to conclusively classify individuals as susceptible, infectious, or recovered from clinically available data.  However, epidemiological process models can be important tools for monitoring trends that reflect population-level infection.  Epidemiological process model conventions sometimes represent simplified biological assumptions made to gain the ability to track population-level trends in clinically available data.  For example, models do not always recognize 1) individual hosts may have detectable levels of pathogenic material without being infectious (i.e., viral RNA), or 2) hosts may also be biologically susceptible to infection with low, detectable antibody levels.  Diagnostics can evaluate when model fit might suffer from potentially oversimplified assumptions.

Hierarchical statistical models and extended epidemiological process models provide methods that link several types of disease surveillance data to epidemiological process models.  Hierarchical models can incorporate methods to account for the sensitivity and specificity of laboratory tests \citep{rogan1978,tabak2019,helman2020,habibzadeh2022}.  Methods that use several types of samples to estimate epidemiological process models often specifically analyze aggregate, population-level incidence and seropositivity rates \citep{bhattacharyya2021,subramanian2021}.  Methods that analyze paired data (data that include multiple sample types from each individual) are rarer.  Paired infection data methods often require a simplifying assumption that the paired data exactly identifies the compartment each individual belongs to \citep{nielsen2007}.  Such assumptions cannot be met when an individual's infection status is not known exactly---e.g., when diagnostic tests do not have perfect sensitivity and specificity, and when results of diagnostic tests do not provide mutually exclusive status (i.e., positive for pathogen and specific antibodies).

Our work builds on previous advances for joint inference of pathogen detection and serology surveillance data by providing general methodology to link paired samples to epidemiological process models.  We specify a hierarchical statistical model that uses an extended SIR model to account for an additional compartment that arises from jointly analyzing several types of surveillance data.  We discuss a theoretical basis for improvement, noting that simpler SIR models may necessarily misclassify individuals as infectious for longer than clinical evidence supports.  The statistical model also accounts for test sensitivity and specificity.  The framework introduces general concepts to facilitate future adaptations to more complicated epidemiological process models and probabilistic sampling designs.  We also propose diagnostic tools to assess model fit.  

We use simulation to demonstrate model properties and apply the model to SARS-CoV-2 surveillance data for North American white-tailed deer (\textit{Odocoileus virginianus}; WTD). National-scale surveillance programs revealed that SARS-CoV-2 emerged in WTD across the United States rapidly, and were detectable at relatively high prevalence ($<35.2\%$) at broad spatial scales---especially for a wildlife pathogen \citep{chandler2021,bevins2023,hewitt2023}. However, the mechanism driving this widespread spatial emergence remain unclear. Developing methods for joint inference of multiple types of surveillance data is an important step towards quantifying the emergence process.  The simulation study demonstrates improvements relative to simpler, conventional SIR modeling.  Simulation also helps identify sample sizes and designs that yield precise estimates for model parameters, similar in some regards to previous studies \citep{pepin2017}.  We use the proposed model to estimate epidemiological dynamics of SARS-CoV-2 in local WTD populations.  The applied analysis highlights how a simple SIR model may overestimate pathogen transmissibility and average recovery times.  SIR models and extensions are widely used in epidemiological modeling, having broad application beyond disease ecology, for example, to public health monitoring.  Our discussion of data integration needs for epidemiological process models is equally important for other diseases, systems, and diagnostic data.

\section{Materials and Methods}

\subsection{The SIR model and proposed SIBR extension}
\label{sec:compartment_models}

The susceptible, infectious, recovered (SIR) compartment model uses a system of differential equations to govern the rates at which the proportion of susceptible $s(\tau)$, infectious $i(\tau)$, and recovered $r(\tau)$ individuals at time $\tau$ evolve.  The SIR system of differential equations specified via 
\begin{align}
\label{eq:sir_ode}
\begin{split}
    \frac{ds(\tau)}{d\tau} &= -\beta s(\tau) i(\tau), \\
    \frac{di(\tau)}{d\tau} &= \beta s(\tau) i(\tau) - \gamma i(\tau), \\    
    \frac{dr(\tau)}{d\tau} &= \gamma i(\tau),
\end{split}
\end{align}
is parameterized with respect to a transmission rate parameter $\beta>0$ and recovery parameter $\gamma>0$.  The SIR model is not sufficient for modeling diseases in which individuals are intermittently infectious, or for modeling demographic effects on population-level infection over time.  Epidemiologists routinely extend the SIR model \eqref{eq:sir_ode} to accommodate more complicated infection systems and observation processes \citep{martcheva2015,faust2018,acemoglu2021,bhattacharyya2021,subramanian2021}.

We propose a susceptible, infectious, \emph{broadly recovered}, recovered (SIBR) extension of the SIR model to support studies that conduct paired pathogen detection and serological tests for all individuals.  In the SIR framework, susceptible individuals have never been infected.  The SIBR model splits the recovered compartment $r(\cdot)$ into broadly recovered $b(\cdot)$ and fully recovered $r(\cdot)$ sub-compartments.  Individuals enter the broadly recovered compartment $b(\cdot)$ when they form an immune response and are no longer infectious, but a targeted characteristic of the pathogen persists such as presence of live pathogen in body tissues  or shedding of non-infectious pathogen.  Biology, research needs, feasibility, and cost can influence the targeted pathogen characteristic.  Individuals transition to the fully recovered compartment $r(\cdot)$ after they fully neutralize the targeted characteristic of the pathogen.  The SIBR system of differential equations extends \eqref{eq:sir_ode} via 
\begin{align}
\label{eq:sibr_ode}
\begin{split}
    \frac{ds(\tau)}{d\tau} &= -\beta s(\tau) i(\tau), \\
    \frac{di(\tau)}{d\tau} &= \beta s(\tau) i(\tau) - \gamma i(\tau), \\
    \frac{db(\tau)}{d\tau} &= \gamma i(\tau) - \eta b(\tau), \\
    \frac{dr(\tau)}{d\tau} &= \eta b(\tau),
\end{split}
\end{align}
with the introduction of a second recovery parameter $\eta>0$.    If the assumption that broadly-recovered individuals are no longer infectious cannot be met, it may be beneficial to reinterpret the transmission parameter $\beta$, modeling it as the product of transmission probability and contact rate terms. \citep[cf.][Section 2.2.1]{acemoglu2021,martcheva2015}.  Alternatively, the ODE \eqref{eq:sir_ode} could be modified to include additional infection pressure (i.e., $s\rightarrow i$ transitions) from individuals in the broadly recovered compartment $b(\cdot)$.

The SIBR model provides more flexibility than the SIR model.  The SIR model \eqref{eq:sir_ode} is a special case of the SIBR model \eqref{eq:sibr_ode}, but both models have distinct features.  The SIBR model can use the broadly $b(\cdot)$ and fully $r(\cdot)$ recovered compartments to model susceptible individuals, infectious individuals, non-infectious individuals with residual pathogenic material, and fully recovered individuals.  By comparison, the SIR model would not be able to model non-infectious individuals with residual pathogenic material separately from fully recovered individuals.

The SIBR model's additional flexibility can more precisely represent epidemiological characteristics of SARS-CoV-2 in WTD and other systems in which individuals can retain viral RNA for several weeks after they are no longer infectious \citep{joynt2020,palmer2021,ramakrishnan2021,griffin2022,martins2022}.  In general, it may be challenging to identify disease systems and diagnostics that satisfy the SIBR model's key assumptions that animals in the broadly recovered state are no longer infectious, and immunity does not wane.  In particular, for SARS-CoV-2 in WTD, the assumptions can be satisfied since the time when neutralizing antibodies are easily detectable via serology corresponds well with the time when animals stop shedding infectious virus \citep{palmer2021}.  Neutralizing antibodies are also observed to persist in WTD for at least 13 months \citep{hamer2022}.  Extensions of the SIBR model that would allow for re-infection may only become important for modeling long-term dynamics of SARS-CoV-2 in WTD, which will also be impacted by population demographics and other factors that routine extensions of disease models accommodate.

\subsection{Characterization maps relate paired surveillance data to models}
\label{sec:characterization_maps}

Individual-level disease surveillance data characterizes individuals as positive or negative for one or more features that inform epidemiological compartment models, such as those discussed in Section \ref{sec:compartment_models}.  For example, serological assays test whether an individual has neutralizing antibodies (NAb) for a pathogen, providing evidence that an individual has broadly or fully recovered from an infection.  Pathogen detection tests such as polymerase chain reaction (PCR) tests and pathogen cultures evaluate whether a host has evidence of the pathogen itself.  Surveillance programs generate statistically \emph{paired} data when they analyze multiple samples or test results for each individual.

Paired data can more precisely associate individuals with specific compartments of epidemiological models.  For example, an individual's viral RNA load and NAb titers can associate individuals with exact compartments of the SIBR model \eqref{eq:sibr_ode}.  We assume an individual without detectable levels of viral RNA or NAbs is susceptible to infection within the SIBR model \eqref{eq:sibr_ode}.  Similarly, we assume an individual with detectable levels of viral RNA but no NAbs is infectious, an individual with detectable levels of viral RNA and NAbs is broadly recovered, and an individual with detectable levels of NAbs but no viral RNA is recovered.  The rules are summarized via the matrix
\begin{align}   
\label{eq:sibr_assignment}
    \begin{blockarray}{cccc}
        & & \BAmulticolumn{2}{c}{T_2} & \\
        & & 0 & 1 \\ \cline{3-4}
        \begin{block}{cc|c|c|}
            \multirow{2}{*}{$T_1$} & 0 & \tS & \tR  \\ \cline{3-4}
                & 1 & \tI & \tB \\ \cline{3-4}
        \end{block}
    \end{blockarray}
\end{align}
in which $T_1=1$ if an individual has detectable levels of viral RNA and $T_1=0$ otherwise, $T_2=1$ if an individual has detectable levels of NAbs and $T_2=0$ otherwise; and $\tS$, $\tI$, $\tB$, and $\tR$ denote that an individual belongs to the susceptible, infectious, broadly recovered, and recovered compartments, respectively.  

In contrast with the SIBR model, an individual's viral RNA load and NAb titers can only partially associate individuals with individual SIR model compartments \eqref{eq:sir_ode}.  There will be some modeling ambiguity because there are four possible outcomes for the paired data $(T_1,T_2)$ but only three compartments in the SIR model.  At least two of the four possible paired data outcomes must be associated with the same compartment within the SIR model.  Drawing on discussion for the SIBR model, analogous SIR associations are summarized via the matrix
\begin{align}
\label{eq:sir_assignment}
    \begin{blockarray}{cccc}
        & & \BAmulticolumn{2}{c}{T_2} & \\
        & & 0 & 1 \\ \cline{3-4}
        \begin{block}{cc|c|c|}
            \multirow{2}{*}{$T_1$} & 0 & \tS & \tR  \\ \cline{3-4}
                & 1 & \tI & \tI/\tR \\ \cline{3-4}
        \end{block}
    \end{blockarray}
\end{align}
in which $\tI/\tR$ highlights the modeling ambiguity.  As discussed in Section \ref{sec:compartment_models}, modelers should draw on \emph{a priori} information about a disease system to associate $(T_1,T_2)=(1,1)$ with compartment $\tI$ or $\tR$.

Association rules can be formulated for more complicated epidemiological compartment models and paired data (i.e., for surveillance programs that run $J>2$ tests for each individual).  In general, surveillance programs can use $J$ features to associate each individual to one of $C$ compartments within an epidemiological compartment model.  The notation $T_j=1$ indicates an individual has the $j$th feature and $T_j=0$ indicates otherwise, where  $j\in\{1,\dots,J\}$.  We formally define a characterization map $\chi:\mT\rightarrow\mC$ to relate paired data outcomes $(T_1,\dots,T_J)\in\mT=\{0,1\}^J$ to an individual's associated epidemiological compartment $Y\in\mC=\{1,\dots,C\}$.  The matrices \eqref{eq:sibr_assignment} and \eqref{eq:sir_assignment} visualize two example characterization maps.  

\subsection{Test positivity observation model}
\label{sec:observation_model}

An individual's true status $T_j$ for feature $j$ may differ from its observed status $T_j^*$ derived from tests due to sample collection challenges and test errors.  Applied analyses that estimate the population-level distribution of infection $\bpi=(\pi_1,\dots,\pi_C)$ and parameters for compartment models such as \eqref{eq:sir_ode} and \eqref{eq:sibr_ode} relate the vector of paired test results $\bT^*=(T_1^*,\dots,T_J^*)$ from each individual to $\bpi$.  In practice, the distribution $\boldsymbol\pi$ will be time-varying for compartment models such as \eqref{eq:sir_ode} and \eqref{eq:sibr_ode}.  The distribution $\boldsymbol\pi$ is also typically modeled hierarchically and will, for example, depend on covariates or spatial features that moderate infection transmission \citep{odea2014,lo2016,faust2018,hewitt2023}.  The population-level distribution of infection has named components $\bpi = (s,i,b,r)$ for the SIBR model \eqref{eq:sibr_ode}.

Test results $\bT^*$ depend on test sensitivity and specificity.   The sensitivity $\varphi_j\in[0,1]$ for test $j$ quantifies the probability that a positive sample yields a positive test result, i.e., $P(T_j^*=1|T_j=1)=\varphi_j$. The specificity $\phi_j\in[0,1]$ quantifies the probability that a negative sample yields a negative test result, i.e.,  $P(T_j^*=0|T_j=0)=\phi_j$.  Test results also depend on a surveillance program's sampling design.  We assume for simplicity that individuals are randomly sampled from the target population.  We also assume the sample size is small relative to the population, to interpret $\pi_y$ as the probability that a randomly selected individual $Y$ is associated with epidemiological compartment $y$.  The marginal observation distribution specified via
\begin{align}
\label{eq:marginal_observation_distribution}
    P(\bT^*=\bt^*\vert\bpi) = \sum_{y\in\mC} 
    P(\bT^*=\bt^*\vert Y = y) \pi_y
\end{align}
links $\bT^*$ to $\bpi$ while accounting for test errors and sampling design.  Joint conditional observation distribution $P(\bT^*=\bt^* |Y=y)$ formulas are straightforward to derive from the characterization map (Supplement, Section 1).

\subsection{Inference}

Disease surveillance programs can combine data from $H$ individuals to estimate parameters for epidemiological compartment models.  The parameter vector $\btheta$ depends on the epidemiological compartment model; for the SIBR model, $\btheta=(\beta,\gamma,\eta)$.  We assume conditional independence between samples and test results, which yields the likelihood specified via 
\begin{align}
    \label{eq:likelihood}
    \ell(\btheta) = \prod_{h=1}^H f(\bt_h^*\vert\btheta,\tau_h),
\end{align}
in which $f(\bt_h^*\vert\btheta,\tau_h)$ is the probability mass function for \eqref{eq:marginal_observation_distribution}.  The distribution \eqref{eq:marginal_observation_distribution} requires the population-level distribution of infection $\bpi(\tau_h;\btheta)$ at the time $\tau_h$ sample $h$ was collected, which is a deterministic function of $\btheta$ and $\tau_h$ for models such as \eqref{eq:sir_ode} and \eqref{eq:sibr_ode}.  

The likelihood \eqref{eq:likelihood} admits both frequentist and Bayesian estimators for $\btheta$ via standard techniques.  Direct optimization of the likelihood \eqref{eq:likelihood} yields a frequentist maximum likelihood estimator $\hat\btheta$ for $\btheta$  and transformations, such as the complete epidemic trajectory $\bpi(\tau;\btheta)$ for arbitrary times $\tau$.  Similarly, specifying a prior distribution for $\btheta$ yields a Bayesian posterior distribution for the same quantities.  When sample sizes are small, Bayesian methods can more reliably quantify uncertainty relative to frequentist methods derived from asymptotic arguments \citep{agresti1998}.

For Bayesian estimators, we specify independent Gamma distributions as priors for $\gamma$, $\eta$, and $R_0=\beta/\gamma$.  Independent Dirichlet and normal distributions can serve as priors for the initial population-level distribution of infection $\bpi(\tau_0;\btheta)$ and outbreak time $\tau_0$ if the parameters are unknown, respectively.  The joint prior distribution for $\gamma$ and $R_0$ induces a prior distribution for $\beta$.  Prior distributions for $\gamma$, $\eta$, and $R_0$ can be simple to parameterize from published estimates of infectiousness and recovery times in the epidemiological literature.  For example, $R_0$ is routinely estimated in epidemiological studies \citep{boonpatcharanon2022}, and a range of plausible $R_0$ values can establish a weakly informative prior for $R_0$ in novel infection systems.  Estimates for typical infection recovery times can be similarly used to establish prior distributions for $\gamma$ and $\eta$.

\subsection{Diagnostics and fit}

We use Bayesian posterior predictive checks to assess model fit for interpretation.  Posterior predictive checks compare simulated data from the posterior distribution to observations \citep[Section 6.3]{gelman2020}.  Posterior predictive sampling requires a posterior sample of $N$ parameter values $\btheta^{(1)},\dots,\btheta^{(N)}$, often obtained via Markov Chain Monte Carlo (MCMC) methods.  Each posterior sample $\btheta^{(n)}$ is used to simulate a replicate dataset from the likelihood \eqref{eq:likelihood}.  Each simulation combines the time-varying population-level distribution of infection $\bpi(\tau;\btheta^{(n)})$ with the observation model \eqref{eq:marginal_observation_distribution} to simulate a dataset ${\bT_1^*}^{(n)},\dots,{\bT_H^*}^{(n)}$ that we compare to the observed surveillance data $\bt_1^*,\dots,\bt_H^*$.  If the model fits the data well, then trends in the time-varying distributions of observed PCR- and sero- positivity among the simulated datasets will resemble trends in the observations.

We use the logarithmic score to assess model fit for model selection.  The log-score is a common choice for binomial response variables \citep{gneiting2007}.  The log-score for model $\mcM$ is specified via
\begin{align*}
    \mcS\left(\mcM\right) = \sum_{h=1}^H\sum_{j=1}^J 
        \log f(t_{hj}^* \vert \hat p_{hj}),
\end{align*}
where $t_{hj}^*$ is the $h$th individual's $j$th test result and $f(\cdot\vert \hat p_{hj})$ is the Bernoulli probability mass function evaluated with estimated success probability $\hat p_{hj}$.  The estimated success probability $\hat p_{hj}$ is the posterior mean for the modeled probability that the test result $t_{hj}^*$ is positive.  Higher log-scores indicate better model fit.

\subsection{Data and designs}

\subsubsection{Simulation study}

\begin{figure}
    \includegraphics[width=\textwidth]{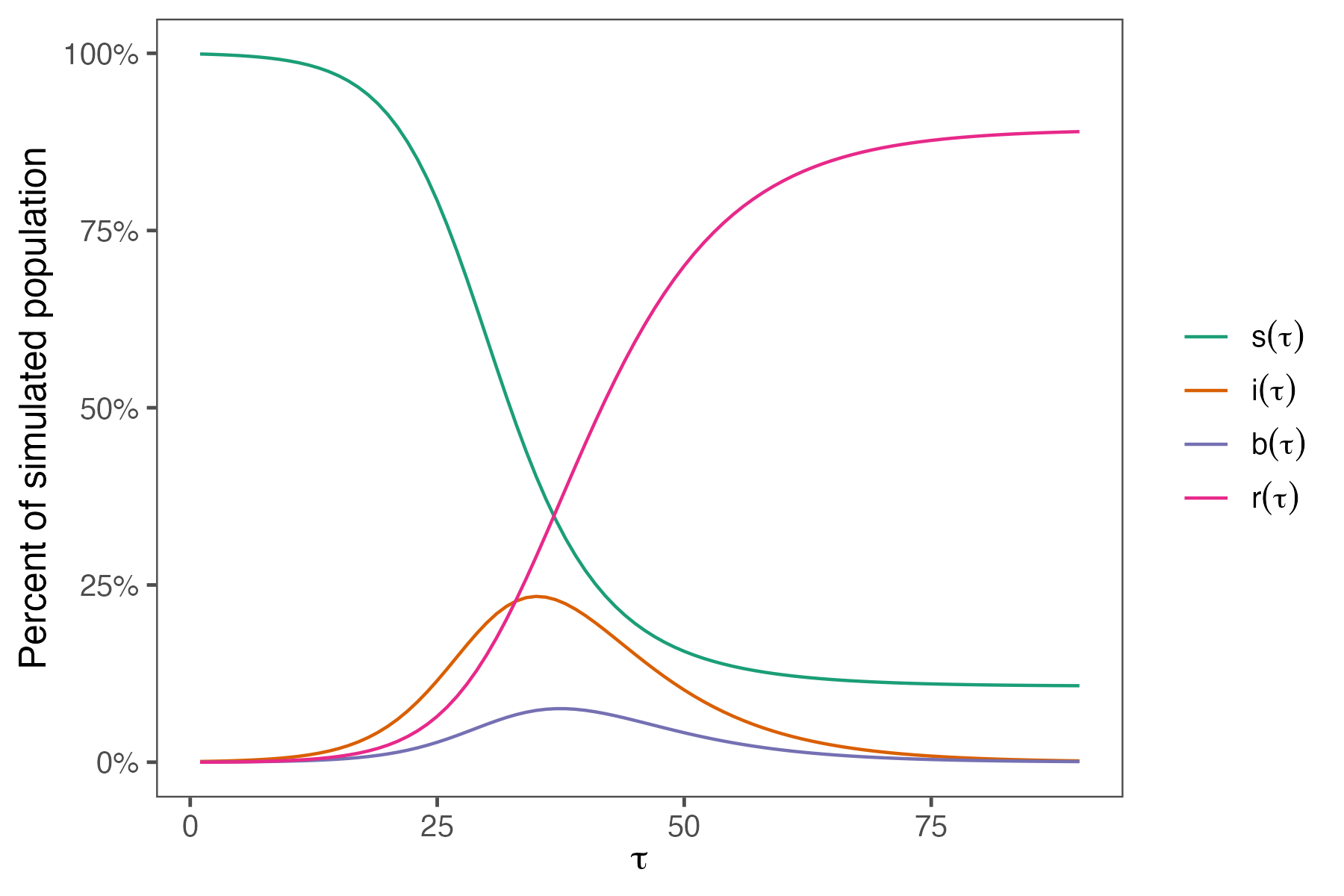}
    \caption{SIBR epidemic curves the simulation study uses, in which time $\tau$ represents days since the start of the simulated epidemic.}
    \label{fig:simulated_trajectory}
\end{figure}

We use simulation to evaluate parameter estimation for the SIBR model under different study designs.  The number of samples and when they are collected vary between designs, in addition to whether the likelihood \eqref{eq:likelihood} uses paired data (i.e., if $J>1$).  The simulated epidemic is the solution to the SIBR differential equations \eqref{eq:sibr_ode} with $\beta=.357$, $\gamma=.143$, $\eta=.429$, and the model is initialized with $s(0)=.999$, $i(0)=.001$, and $b(0)=r(0)=0$.  The model has reproductive number $R_0=\beta/\gamma=2.50$, maximum infectious proportion of .23 at 34 days, and takes 95 days before infectious levels fall below $.001$ (Fig.\ \ref{fig:simulated_trajectory}). 

We simulate paired data for test results that can identify viral RNA (e.g., PCR) and NAbs (e.g., serology).  The simulation's sensitivity and specificity is perfect---i.e., $\phi_{1}=\phi_{2}=\varphi_{1}=\varphi_{2}=1$.  The simulated data uses ideal test conditions to evaluate the impacts of study design on its own.  Previous studies investigated the impact of sensitivity and specificity on epidemic predictions \citep{bhattacharyya2021}.

We use the simulated epidemic to analyze 1,000 replicate datasets for each of 70 study designs.  Designs sample individuals from one random day within each monthly, biweekly, or weekly time period of the simulated epidemic.  For example, designs with a monthly sampling rate will sample individuals from 3 random days of the simulated epidemic with the first date between day 1--30, the second date between day 31--60, and the last date between day 61--90.  Designs sample a total of either 10, 50, 100, 200, or 500 individuals.  We do not evaluate study designs with 10 individuals and weekly sampling since this would require at least 13 samples.  Supplement \simconfigurations{} explicitly shows the sample size and frequency combinations.  The number of individuals sampled from each time period are either evenly divided or randomly distributed.  The random distributions are either uniform or biased toward early, middle, or late time periods (Supplement, \simsamplesizefig{}).  Each sampled individual is randomly assigned to one of the SIBR compartments using the epidemic curve as compartment weights (i.e., with randomness following Fig.\ \ref{fig:simulated_trajectory}).  Study designs also differ between whether the simulated study only conducts PCR tests, serology tests, or both types of tests.  The association rules \eqref{eq:sibr_assignment} determine the individual's corresponding test results.  For example, simulated susceptible individuals will have negative PCR and serology tests.  

The simulation study uses Bayesian posteriors to estimate parameters for simulated data since the simulation explores datasets with small sample sizes.  We specify a diffuse Gamma prior distribution for $R_0$ with mean and variance set to 2.5 and 100, respectively.  We specify diffuse Gamma prior distributions for $\gamma$ and $\eta$ as well, each with mean and variance both set to 2.  We estimate the initial outbreak time $\tau_0$, specifying a mean-zero normal prior distribution with variance set to 100.  The initial conditions $s(\tau_0)$, $i(\tau_0)$, $b(\tau_0)$, $r(\tau_0)$ are treated as known, which is reasonable for surveillance programs that continuously monitor for outbreaks.  We only estimate parameters for the SIBR model to avoid biased study design comparisons from fitting explicitly misspecified models (i.e., from estimating SIR-I parameters for data generated from an SIBR model).

We use the continuous ranked probability score (CRPS) to assess how well each study design recovers model parameters; better designs yield smaller scores \citep{gneiting2007}. The CRPS generalizes mean absolute error (MAE) to summarize both the accuracy and precision of the posterior distribution for each parameter.  The CRPS is small (i.e., less than 10\% of the parameter's value) when 1) the posterior distribution's median is close to the true parameter, and 2) the posterior distribution is concentrated (i.e., as variance decreases).  The posterior median is a natural target to study because the simulation is non-Gaussian and includes some small datasets, which naturally yield skewed posterior distributions.  The posterior mean and mean square error are more appropriate summaries for symmetric posterior distributions that arise from analyzing larger datasets or models with Gaussian likelihoods, for example.

\subsubsection{Case study data: SARS-CoV-2 in White-tailed deer}

We use data from a national surveillance study of SARS-CoV-2 in WTD \citep{bevins2023} to demonstrate our method of joint analysis of PCR and serology data.  WTD samples were collected postmortem from multiple sources, including hunter harvest samples collected by state departments of natural resources, management events conducted by United States Department of Agriculture Animal and Plant Health Inspection Service (USDA-APHIS), Wildlife Services, and opportunistic sampling of mortalities such as roadkill collected by all agencies.  Nasal or oral swabs were collected and tested for the presence of SARS-CoV-2 viral RNA via real-time reverse-transcriptase polymerase-chain-reaction  (rRT-PCR) as described in \citet{bevins2023}.  Whole blood samples were collected on Nobuto filter paper and the presence of SARS-CoV-2 antibodies was inferred via a surrogate virus neutralization test \citep{bevins2023}.

We analyze a subset of the national surveillance study that were well sampled during the study's November 2021 to October 2022 sampling period.  We focus on three counties in the northeast United States, each with at least 100 samples that were collected at a minimum rate of once every two weeks for at least ten weeks.  The counties vary between 800--2,000 sq. km. in size.  County A is farther than 350 km from counties B and C.  County A is mostly rural, but contains a metropolitan area with approximately 150,000 residents. County B is mostly suburban, and County C is mostly rural.  Counties B and C share a border, so their deer populations and disease processes may potentially overlap and be dependent.  However, we model the counties separately and do not attempt to draw inferences about cross-site processes that may potentially be influenced via dependence.  More complex hierarchical models can account for spatial dependence between disease processes but require substantially greater disease process replication---i.e., by simultaneously analyzing data from hundreds of counties \citep{irvine2007}.  Hierarchical models that more explicitly consider disease transmission by animal movement could also be appropriate \citep{davis2019}.  Instead, we focus on comparing within-county Bayesian fits to several combinations of models and data.  We assume PCR tests have perfect sensitivity and specificity---i.e., that $\phi_{1}=\varphi_{1}=1$.

The performance of the sVNT assay for WTD Nobuto samples has not been validated through a controlled laboratory study with well characterized positive and negative samples, so the analytical sensitivity and specificity are unknown.  Therefore, we evaluate the reliability of results to different assumptions about potential Nobuto strip test sensitivity and specificity.  First, we assume for the purposes of comparison that test sensitivity is known to be 98.9\% and specificity is 100\%, following sera studies in humans and some companion animals \citep{tan2020,perera2021}.  Wildlife disease studies routinely use test protocols and assumptions that have been validated for other species as a convention, out of necessity, since more relevant data are often difficult to obtain \citep{jia2020}.  Second, we estimate Nobuto strip test sensitivity and specificity relative to Beta prior distributions $\varphi_{2}\sim\text{Beta}(7,6)$ and $\phi_{2}\sim\text{Beta}(41,1)$.  The prior distributions are conjugate posteriors (relative to uninformative Uniform priors) from a separate analysis of 51 free ranging WTD (unpublished data).  Blood was collected from the WTD via venipuncture and Nobuto strips.  The separate analysis treats venipuncture test results as ground truth that Nobuto tests should replicate.  Empirical sensitivity for the Nobuto tests was 55\% ($n=11$) and empirical specificity was 100\% ($n=40$).

\section{Results}

\subsection{Simulation}

\begin{figure}
    \centering
    \includegraphics[width=\textwidth]{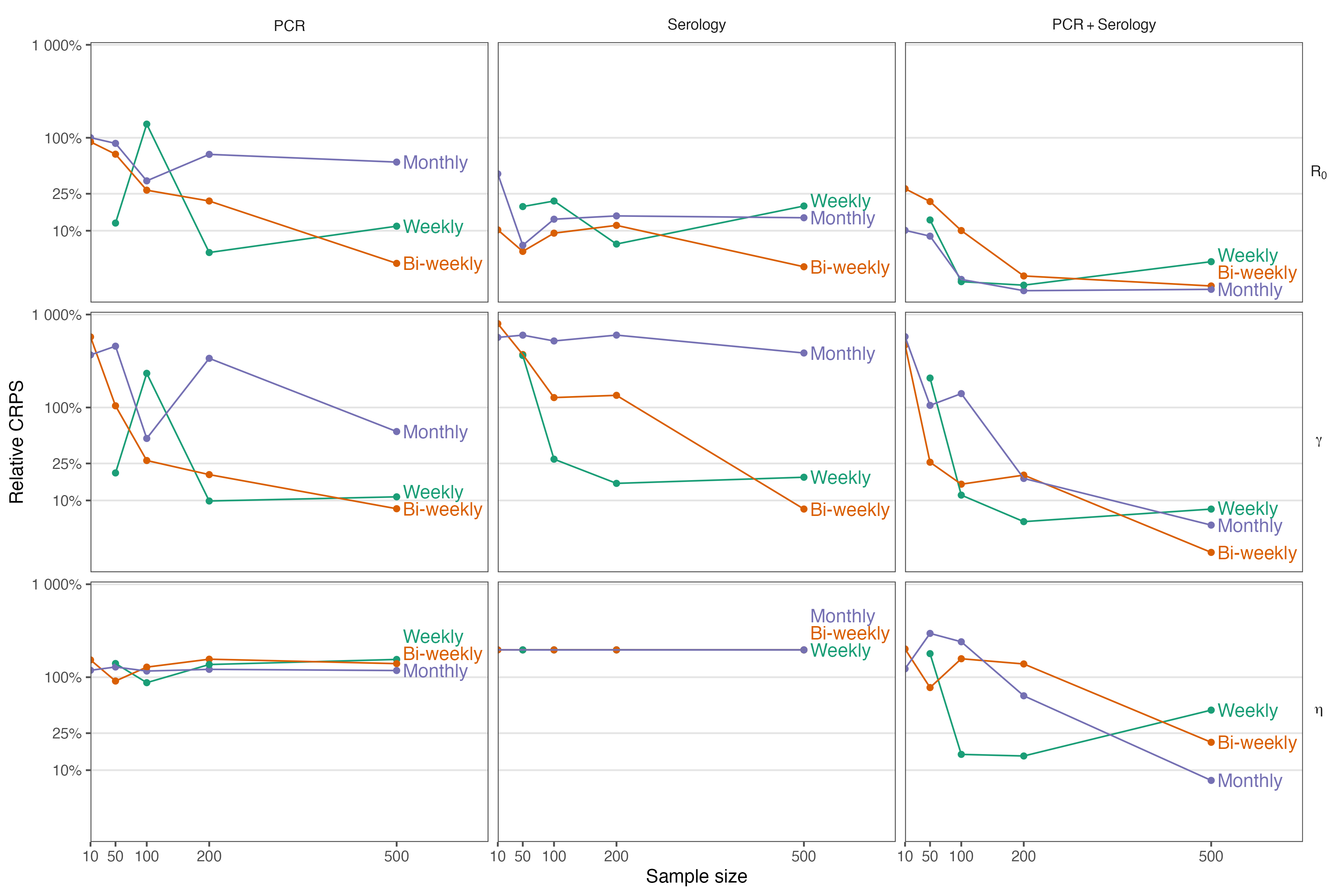}
    \caption{Relative CRPS averaged across parameter estimates from 1,000 simulated datasets and select study designs.  Study designs vary sample size and collection times, but all have a uniformly distributed number of samples taken at each timepoint.  Relative CRPS is the CRPS for the parameter estimate divided by the parameter's true value.}
    \label{fig:sim_accuracy_sample_size}
\end{figure}

Sample size and data (i.e., PCR, serology, or both) influence parameter estimation the most.  The most precise and accurate parameter estimates tend to require at least 100 paired PCR and serology samples collected over time (Fig.\ \ref{fig:sim_accuracy_sample_size}).  Relative CRPS always tends to be smaller for paired data designs than for designs that only use PCR or serology data alone.  The reproduction number $R_0$ requires at least 100 samples to achieve less than 10\% relative CRPS, regardless of how samples are collected over time.  Relative CRPS for the broad recovery rate $\gamma$ is close to 10\% when at least 100 samples are collected at weekly or bi-weekly intervals.  The full recovery rate $\eta$ is difficult to estimate precisely in all study designs.  Relative CRPS for $\eta$ tends to decrease with sample size when both PCR and serology are collected, but not when only one data type, PCR or serology is collected.  The observation indicates that $\eta$ is only estimable from paired data.

\begin{figure}
    \centering
    \includegraphics[width=\textwidth]{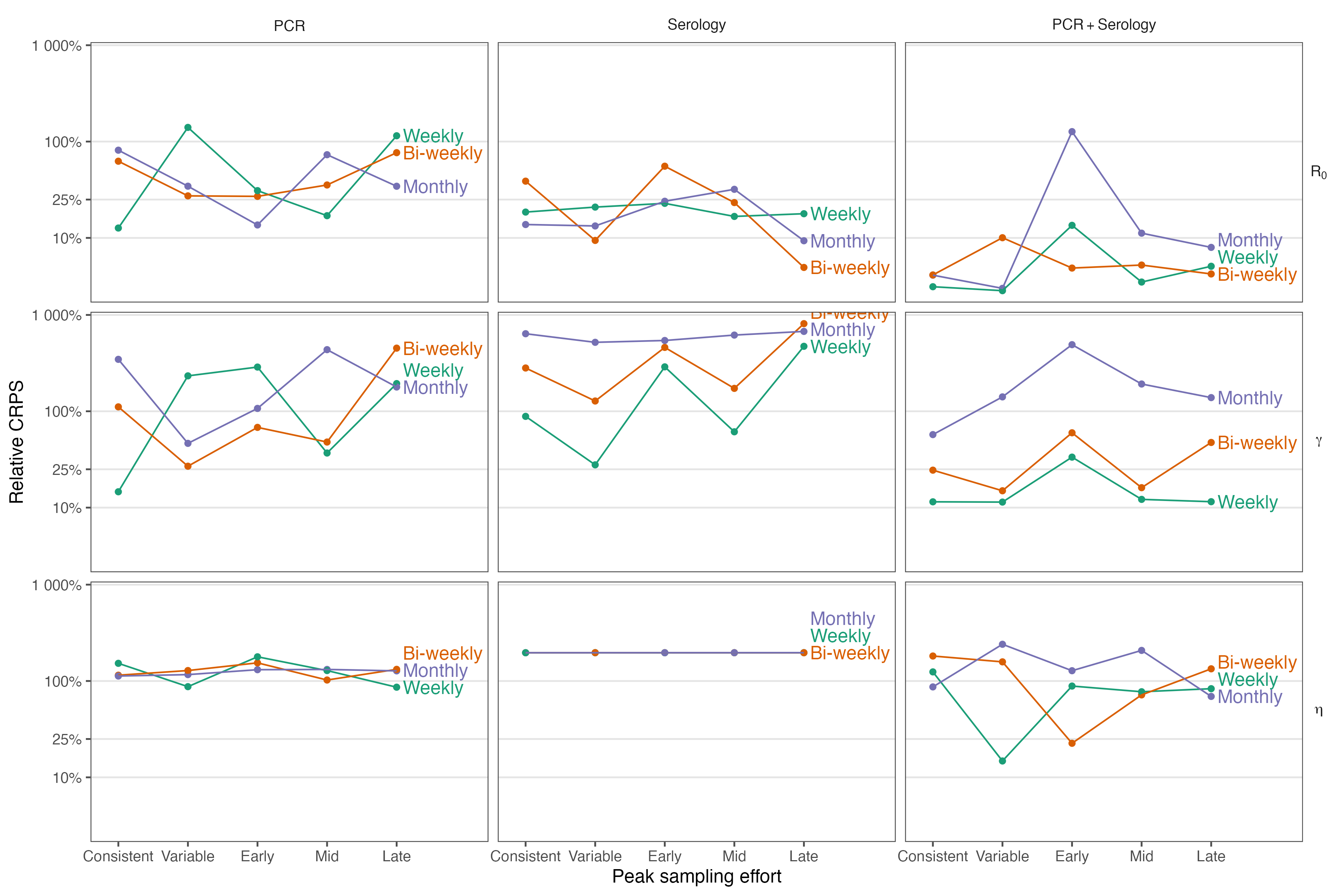}
    \caption{Relative CRPS averaged across parameter estimates from 1,000 simulated datasets and select study designs.  Study designs vary collection times and the numbers of individuals sampled from each time period, but all use a total of 100 samples to estimate parameters.  Relative CRPS is the CRPS for the parameter estimate divided by the parameter's true value.}
    \label{fig:sim_accuracy_timing}
\end{figure}

There are complex tradeoffs between estimator precision and accuracy, how often samples are collected over time, and how many samples are collected from each timepoint.  For each data type and sample collection interval combination (i.e., for each line in Fig.\ \ref{fig:sim_accuracy_timing}), the most precise and accurate parameter estimates tend to come from study designs that collect an equal or uniformly distributed number of samples from each timepoint.  There are some exceptions.  Monthly sampling concentrates sampling efforts into fewer sampling timepoints (i.e., site visits), making parameter estimation more challenging by having fewer opportunities to observe epidemic dynamics over time.  If feasible for field teams, concentrating sampling in either early or late stages of an outbreak (i.e., where infection is primarily spreading or waning) can improve estimates of transmission or recovery parameters, respectively.

\subsection{SARS-CoV-2 in White-tailed deer}

Model comparison suggests analyses should jointly fit the SIBR model to PCR and serology data and estimate sensitivity and specificity for serology data.  The SIBR model fit scores substantially improve when jointly analyzing PCR and serology data, as compared to analyzing data separately (Table \ref{table:model_comparison}).  The SIBR model fit scores also suggest analyses are more sensitive to the choice to include or exclude PCR data than the choice to include or exclude serology data.  Models that treat Nobuto test sensitivity and specificity as known fit the data about as well as models that estimate Nobuto test sensitivity and specificity.  Model fit scores indicate the SIBR model behaves like the SIR-I model when only PCR data is analyzed and like the SIR-R model when only serology data is analyzed.

\begin{table}[ht]
\centering
\caption{Log-score model fit values for different combinations of data at  each site.  Models are fit using the data indicated by the  ``Data'' columns.  The ``Est. sens./spec.'' column indicates  estimating sensitivity and specificity for Nobuto tests.  Log-scores evaluate each model's expected test positivity rates  for different data subsets: PCR data alone  serology data alone, or the combined dataset where possible.  The SIBR model's structure always allows it to compute expected  test positivity rates for PCR and Serology data. By comparison, SIR-R and SIR-I models are evaluated with respect to serology or  PCR data only, respectively, since the simpler models cannot  make inference about the other data type directly.  The best PCR, Serology, and Combined log-scores within each county are bolded.} 
\label{table:model_comparison}
\begin{tabular}{llccclll}
  & & \multicolumn{3}{c}{Data} & \multicolumn{3}{c}{Score}\\
Location & Model & PCR & Serology & Est. sens./spec. & PCR & Serology & Combined \\ 
  \midrule
\multirow{9}{*}{County A} & \multirow{5}{*}{SIBR} & $\bullet$ &  &  & $\boldsymbol{-98}$ & $-167$ & $-265$ \\ 
   &  & $\bullet$ & $\bullet$ &  & $\boldsymbol{-98}$ & $\boldsymbol{-151}$ & $\boldsymbol{-249}$ \\ 
   &  & $\bullet$ & $\bullet$ & $\bullet$ & $\boldsymbol{-98}$ & $\boldsymbol{-151}$ & $\boldsymbol{-249}$ \\ 
   &  &  & $\bullet$ &  & $-349$ & $\boldsymbol{-151}$ & $-500$ \\ 
   &  &  & $\bullet$ & $\bullet$ & $-169$ & $\boldsymbol{-151}$ & $-320$ \\
        \addlinespace[2pt]\cline{2-8}\addlinespace[5pt]
        & \multirow{2}{*}{SIR-R} &  & $\bullet$ &  &  & $\boldsymbol{-151}$ &  \\ 
   &  &  & $\bullet$ & $\bullet$ &  & $\boldsymbol{-151}$ &  \\
        \addlinespace[2pt]\cline{2-8}\addlinespace[5pt]
        & \multirow{2}{*}{SIR-I} & $\bullet$ &  &  & $\boldsymbol{-98}$ &  &  \\ 
   &  & $\bullet$ &  & $\bullet$ & $\boldsymbol{-98}$ &  &  \\ 
   \midrule
\multirow{9}{*}{County B} & \multirow{5}{*}{SIBR} & $\bullet$ &  &  & $\boldsymbol{-26}$ & $-97$ & $-123$ \\ 
   &  & $\bullet$ & $\bullet$ &  & $-29$ & $\boldsymbol{-77}$ & $-106$ \\ 
   &  & $\bullet$ & $\bullet$ & $\bullet$ & $\boldsymbol{-26}$ & $\boldsymbol{-77}$ & $\boldsymbol{-104}$ \\ 
   &  &  & $\bullet$ &  & $-54$ & $-78$ & $-132$ \\ 
   &  &  & $\bullet$ & $\bullet$ & $-61$ & $\boldsymbol{-77}$ & $-139$ \\
        \addlinespace[2pt]\cline{2-8}\addlinespace[5pt]
        & \multirow{2}{*}{SIR-R} &  & $\bullet$ &  &  & $-78$ &  \\ 
   &  &  & $\bullet$ & $\bullet$ &  & $\boldsymbol{-77}$ &  \\
        \addlinespace[2pt]\cline{2-8}\addlinespace[5pt]
        & \multirow{2}{*}{SIR-I} & $\bullet$ &  &  & $\boldsymbol{-26}$ &  &  \\ 
   &  & $\bullet$ &  & $\bullet$ & $\boldsymbol{-26}$ &  &  \\ 
   \midrule
\multirow{9}{*}{County C} & \multirow{5}{*}{SIBR} & $\bullet$ &  &  & $\boldsymbol{-62}$ & $-146$ & $-208$ \\ 
   &  & $\bullet$ & $\bullet$ &  & $-69$ & $-104$ & $\boldsymbol{-173}$ \\ 
   &  & $\bullet$ & $\bullet$ & $\bullet$ & $-71$ & $-104$ & $-175$ \\ 
   &  &  & $\bullet$ &  & $-197$ & $\boldsymbol{-103}$ & $-300$ \\ 
   &  &  & $\bullet$ & $\bullet$ & $-91$ & $-104$ & $-195$ \\
        \addlinespace[2pt]\cline{2-8}\addlinespace[5pt]
        & \multirow{2}{*}{SIR-R} &  & $\bullet$ &  &  & $\boldsymbol{-103}$ &  \\ 
   &  &  & $\bullet$ & $\bullet$ &  & $-104$ &  \\
        \addlinespace[2pt]\cline{2-8}\addlinespace[5pt]
        & \multirow{2}{*}{SIR-I} & $\bullet$ &  &  & $\boldsymbol{-62}$ &  &  \\ 
   &  & $\bullet$ &  & $\bullet$ & $\boldsymbol{-62}$ &  &  \\ 
   \bottomrule
\end{tabular}
\end{table}

\begin{figure}
    \centering
    \includegraphics[width=\textwidth]{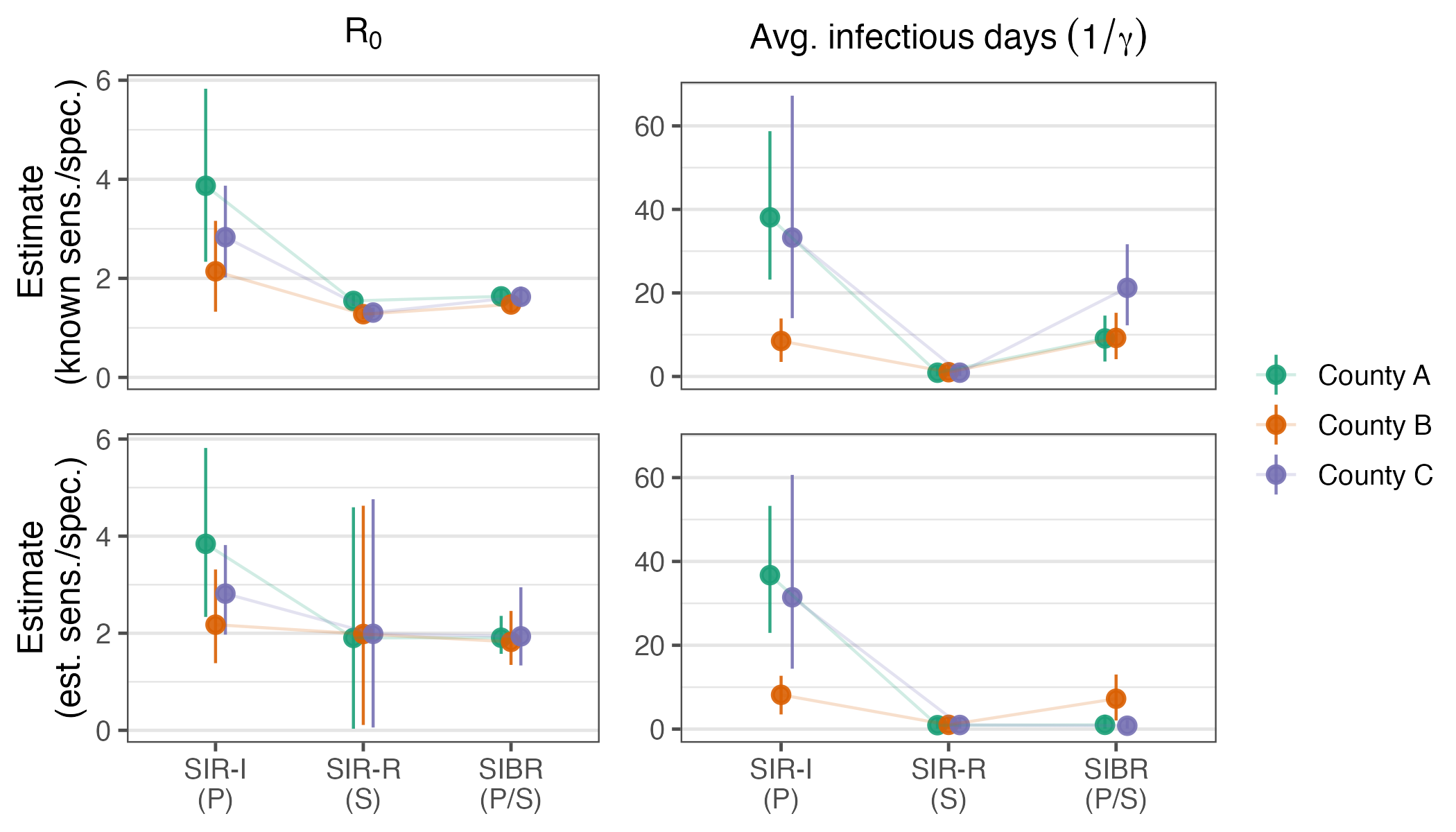}
    \caption{Posterior summaries (mean and 95\% HPDI's) for reproduction numbers $R_0$ and the average number of infectious days ($1/\gamma$) for different counties.  Parameter estimates depend on the model (SIR-I, SIR-R, SIBR) and choice to use PCR (P), Serology (S), or both (P/S) types of data during estimation and whether sensitivity and specificity is estimated for Nobuto tests.}
    \label{fig:parameter_comparison}
\end{figure}

Estimates for epidemiological parameters depend on the model, data subsets used, and choice to treat Nobuto test sensitivity and specificity as known or estimated.  For example, reproductive number $R_0$ estimates based only on PCR data are several times larger than estimates that incorporate serology data (Fig.\ \ref{fig:parameter_comparison}).  Estimates for the average number of days individuals will be infectious are similarly impacted.  The model comparison results suggest the SIBR model jointly fit to PCR and serology data should be preferred (Table \ref{table:model_comparison}).  In further support of the preference, only the SIBR model jointly fit to PCR and serology data provides strong posterior learning for all model parameters relative to the prior distribution (Supplement, \modelfit{}).

\begin{figure}
    \centering
    \includegraphics[width=\textwidth]{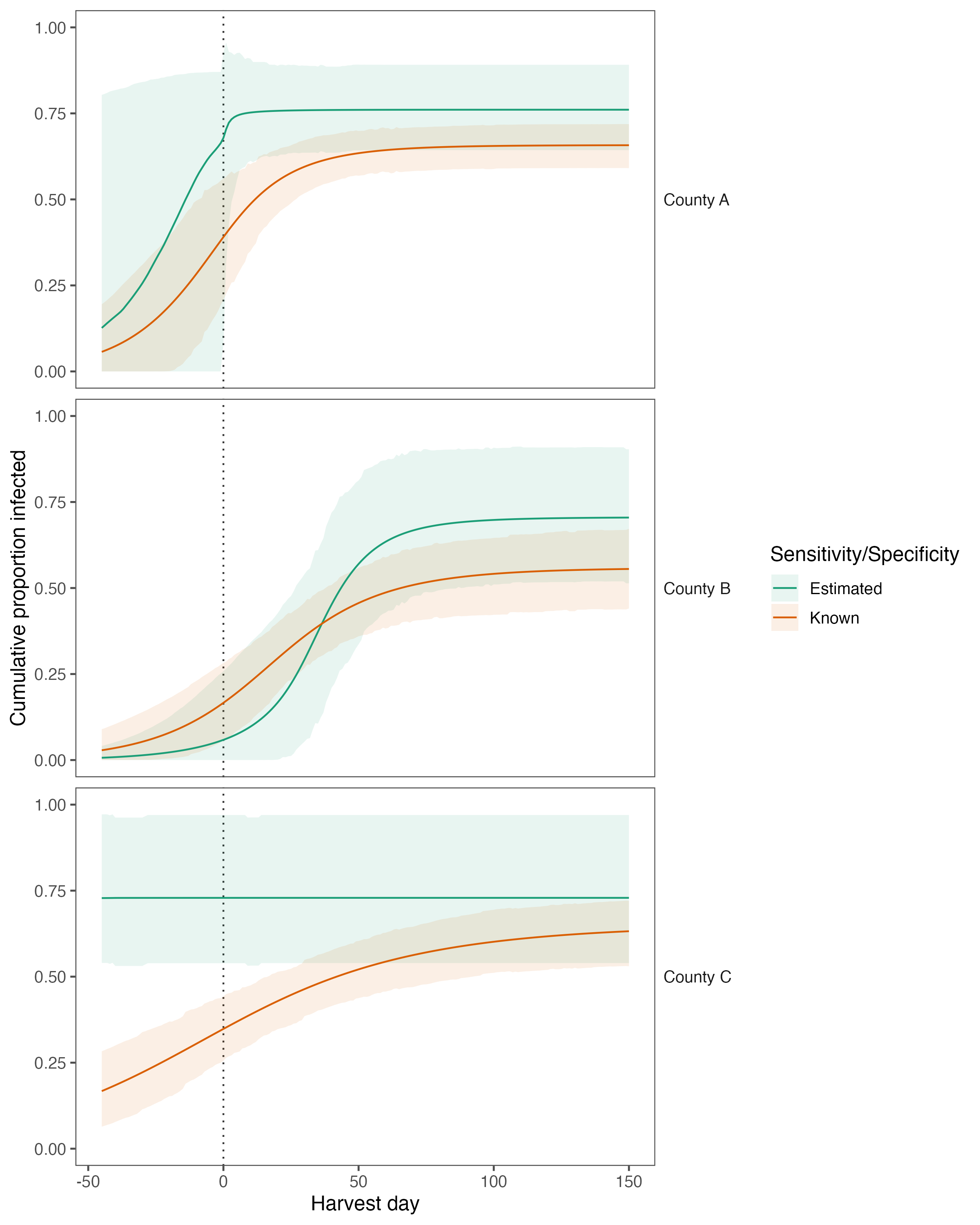}
    \caption{Cumulative proportion of WTD estimated to have been infected.  Estimates are for the SIBR model fit using both PCR and Serology data.  The vertical dotted lines indicate when data collection begins.  Negative dates indicate projections before sampling occurred.}
    \label{fig:epidemic_dynamics}
\end{figure}

Based on model comparison, we focus on interpreting results in more detail for the SIBR model that incorporates sensitivity and specificity estimation and is jointly fit to PCR and serology data.  The model captures trends in the data well and provides estimates of cumulative infections.  To assess model fit, we note that observed PCR- and sero- positivity rates for daily swab data are consistent with the range of rates the fitted model predicts (Supplement, \posteriorlearning{}).  Infection is estimated to spread at similar rates in counties A ($R_0$ est.\ 1.92; 95\% highest posterior density interval (HPDI): 1.55--2.39) and C ($R_0$ est.\ 1.92; 95\% HPDI: 1.35--2.96), and slightly slower in county B ($R_0$ est.\ 1.79; 95\% HPDI: 1.34--2.43).  For each county, infection rates are estimated to be slightly higher when estimating sensitivity and specificity.  Estimates for the average number of infectious days (i.e., $1/\gamma$) are 1.32 days (95\% HPDI: .13--6.34), 7.57 days (95\% HPDI: 2.69--13.50), and 0.80 days (95\% HPDI: .11--2.13) days for counties A--C, respectively.  Estimates for all model parameters are available in the Supplemental material (\countytables{}).  SARS-CoV-2 may have been introduced in county C earlier than county A, leading to lower observed PCR positivity and estimated infection rates during the study period (Fig.\ \ref{fig:epidemic_dynamics} and Supplement, \modelfit{}).  Evidence for earlier introduction in county C is much stronger in models that estimate sensitivity and specificity, which suggest the outbreak was effectively over before sampling began.  The model estimates large proportions of WTD were infected by the end of the study period (April 2023).  At the end of the study period, 76\% (95\% HPDI: 64--89\%), 71\% (95\% HPDI: 51-90\%), and 73\% (95\% HPDI: 54--97\%) of WTD are estimated to have been infected by SARS-CoV-2 in counties A--C, respectively when estimating sensitivity and specificity.  Infection rates are slightly higher when estimating sensitivity and specificity.

\section{Discussion}

We contribute a hierarchical statistical modeling framework and extended SIR model that links paired sample diagnostic results from wildlife disease surveillance programs to epidemiological process models.  Methods that simultaneously analyze all samples recognize that paired data represent more detailed observations of individuals than any one data stream provides individually.  More detailed observations can improve understanding of disease transmission and persistence.  

The use of characterization maps in the theoretical development of the SIBR models provides a theoretical basis to remedy potential biases that can arise by fitting epidemiological process models without paired data. For example, the SIR model has the potential to overestimate infectiousness of SARS-CoV-2 in WTD if only PCR data are available since infectiousness is less precisely known.  We demonstrate empirically and through simulation that developing paired data methods for epidemiological process models can increase the precision for epidemiological parameter estimates.  Paired data methods have the potential for broad application since many wildlife surveillance programs already collect several types of samples from each individual.

We use the SIBR model to estimate epidemiological parameters for SARS-CoV-2 in wild WTD.  It is the first paired data analysis for SARS-CoV-2 epidemiological dynamics in wild WTD.  Bayesian paired data methods regularly improve disease detection for individual humans but have not necessarily been used with diagnostic test data to estimate epidemiological dynamics \citep{berkvens2006,dendukuri2001,cao2021}.  More recently, paired data methods are being developed to integrate infection occurrence data with specialized datasets, such as genetic sequence data \citep{volz2009,lau2015,smith2017,volz2018,featherstone2022,king2022} and movement data \citep{andrade2022,fox2022}.

The SIBR model with estimated Nobuto test sensitivity and specificity suggests the average time WTD can remain infectious for SARS-CoV-2 may be up to 2--14 days via 95\% HPDI upper bounds.  The relatively short infectious periods correspond to captive animal studies that show live virus cannot be isolated after 5--7 days \citep{palmer2021,martins2022}.  Infectiousness may potentially be longer in some wild WTD populations than captive populations, or only appear so due to differing deer-to-deer contact networks.  Infectiousness estimates, linked to $R_0$, incorporate both deer-to-deer transmission and human-to-deer spillover.  Studies that formally incorporate phylogenetic data into epidemiological process models can potentially help separate different contributions to overall transmission \citep[cf.][]{featherstone2022}.

By comparison, the SIBR model that assumes Nobuto test sensitivity and specificity are known suggests the average time WTD can remain infectious may be much longer, with 95\% HPDI upper bounds ranging up to 32 days.  The apparent discrepancy in analytic results is directly due to the Nobuto strip sensitivity and specificity assumption.  The much longer infectious time seems biologically unlikely given captive animal study results, which suggests samples collected on Nobuto strips do not perform as well as samples directly extracted from sera.  Our model estimates sensitivity and specificity for Nobuto strips.  However, testing additional paired serological samples directly extracted from sera and collected on Nobuto strips can help improve estimates of Nobuto strip sensitivity and specificity.

The SIBR model can be embedded in hierarchical models that estimate landscape-scale patterns in epidemiological parameters.  In general, landscape-scale models can account for spatial autocorrelation and the influence of environmental covariates \citep{meentemeyer2012,caprarelli2014,pepin2019,podgorski2020}.  Hierarchical models pool data across space and thus can provide more precise parameter estimates than single-site analyses and predict at unobserved locations.  Data pooling requires an assumption that parallel epidemiological dynamics occur in nearby locations.  By jointly analyzing PCR and serological data, SIBR estimates for $R_0$ already have equal or smaller posterior uncertainty than those presented in a landscape-scale analysis of SARS-CoV-2 in WTD based on SIR processes fit to PCR data \citep{hewitt2023}.  Embedding the SIBR model in landscape-scale models can potentially further reduce uncertainty via data pooling.  

Surveillance programs can use the simulation study to plan site-level sampling designs for the SIBR model.  The simulation study suggests epidemiological dynamics parameters are best learned from a total of at least 100 paired PCR and serological samples from weekly or biweekly site visits.  The study also suggests the full recovery parameter $\eta$ requires paired data to estimate.  The requirement is reasonable since transitions into the fully recovered epidemiological compartment $R$ is only identifiable with paired data.  Our recommended sample sizes are smaller than findings from other simulation studies of epidemiological parameter estimation \citep{vinh2015}.  We attribute the difference to the use of paired data.  The SIBR model adds value over a simpler, SIR model when surveillance programs can collect time-varying PCR and serological data.  When only PCR-positive or seropositive samples can be collected (but not both), we demonstrate the SIBR model is equivalent to a simpler SIR model since the SIBR model directly extends the SIR model.  The simulation study we present is only informative for sampling designs for infectious diseases with similar dynamics as SARS-CoV-2 in WTD.  The study would need to be modified to evaluate sampling plans for infections with, for example, substantially longer or shorter infectious or recovery periods, or for different species.

We use general statistical theory to develop methods to combine paired samples and motivate an extended SIR model. The general theory is applicable to more complex process models that, for example, allow for reinfection or demographic shifts across years \citep[see][for review]{martcheva2015}.  However, more complex epidemiological processes could potentially require more than two types of samples from each animal, and more than one additional epidemiological compartment to identify infection status or history.  The characterization maps, observation distributions, and other concepts we specify use notation that directly supports such extensions. 

We contribute methods to better estimate epidemiological process models when paired data is sampled from individuals via wildlife surveillance programs.  Many wildlife surveillance programs already collect several samples from individuals and could potentially apply or adapt the methods we propose.  Wildlife surveillance programs often have spatial scales or ecological or budget constraints that limit the ability to implement robust probabilistic sampling designs.  Methods that can more efficiently learn epidemiological processes from relatively limited amounts of data can improve wildlife surveillance program efficacy.


\section{Conflicts of interest}

The authors declare there are no conflicts of interest for this paper.

\section{Acknowledgements}

We thank the federal employees at USDA APHIS Wildlife Services, USDA APHIS National Wildlife Disease Program, and collaborators at state wildlife agencies for contributing wildlife sampling expertise, as well as hunters for participating in this effort.  We also thank Sarah Bevins, Jonathon Heale, Allen Gosser, Richard Chipman, Julianna Lenoch, and John Short for useful discussion and WTD program coordination.  This study was funded by the American Rescue Plan Act through the U.S. Department of Agriculture, Animal and Plant Health Inspection Service.  MCMC samplers implemented using NIMBLE v0.13.1 \citep{nimble-software}.  We also thank two anonymous reviewers for their feedback.

\section{Author contributions}

JH, GWH, and KP conceived the ideas and designed methodology; DC, JR, CQ, RP, JC, SB, and TJD collected and helped interpret the data; JH and GWH analyzed the data; JH led the writing of the manuscript. All authors contributed critically to the drafts and gave final approval for publication.


\bibliography{references}

\begin{thebibliography}{}

\bibitem[Acemoglu et~al., 2021]{acemoglu2021}
Acemoglu, D., Fallah, A., Giometto, A., Huttenlocher, D., Ozdaglar, A., Parise,
  F., and Pattathil, S. (2021).
\newblock Optimal adaptive testing for epidemic control: combining molecular
  and serology tests.
\newblock {\em arXiv preprint arXiv:2101.00773}.

\bibitem[Agresti and Coull, 1998]{agresti1998}
Agresti, A. and Coull, B.~A. (1998).
\newblock Approximate is better than “exact” for interval estimation of
  binomial proportions.
\newblock {\em The American Statistician}, 52(2):119--126.

\bibitem[Anderson and May, 1991]{anderson1991}
Anderson, R.~M. and May, R.~M. (1991).
\newblock {\em Infectious diseases of humans: dynamics and control}.
\newblock Oxford university press.

\bibitem[Andrade and Duggan, 2022]{andrade2022}
Andrade, J. and Duggan, J. (2022).
\newblock Inferring the effective reproductive number from deterministic and
  semi-deterministic compartmental models using incidence and mobility data.
\newblock {\em PLoS Computational Biology}, 18(6):e1010206.

\bibitem[Berkvens et~al., 2006]{berkvens2006}
Berkvens, D., Speybroeck, N., Praet, N., Adel, A., and Lesaffre, E. (2006).
\newblock Estimating disease prevalence in a {Bayesian} framework using
  probabilistic constraints.
\newblock {\em Epidemiology}, 17(2):145--153.

\bibitem[Bevins et~al., 2021]{bevins2021}
Bevins, S.~N., Chandler, J.~C., Barrett, N., Schmit, B.~S., Wiscomb, G.~W., and
  Shriner, S.~A. (2021).
\newblock Plague exposure in mammalian wildlife across the western {United}
  {States}.
\newblock {\em Vector-Borne and Zoonotic Diseases}, 21(9):667--674.

\bibitem[Bevins et~al., 2023]{bevins2023}
Bevins, S.~N., Chandler, J.~C., Beckerman, S., Bergman, D.~L., Chipman, R.~B.,
  Collins, D.~T., Eckery, J.~P., Ellis, J.~W., Gosser, A.~L., Heale, J.~D.,
  Klemm, J., Lantz, K., Lindera, T.~J., Pleszewski, R., Quintanal, C.,
  Ringenberg, J., Weir, K.~R., Torchetti, M.~K., Lenoch, J.~B., DeLiberto,
  T.~J., and Shriner, S.~A. (2023).
\newblock {SARS-CoV-2} occurrence in white-tailed deer throughout their range
  in the continental {United} {States}.
\newblock {\em bioRxiv}, 2023.04.14.533542.

\bibitem[Bhattacharyya et~al., 2021]{bhattacharyya2021}
Bhattacharyya, R., Kundu, R., Bhaduri, R., Ray, D., Beesley, L.~J., Salvatore,
  M., and Mukherjee, B. (2021).
\newblock Incorporating false negative tests in epidemiological models for
  {SARS-CoV-2} transmission and reconciling with seroprevalence estimates.
\newblock {\em Scientific Reports}, 11:9748.

\bibitem[Boonpatcharanon et~al., 2022]{boonpatcharanon2022}
Boonpatcharanon, S., Heffernan, J.~M., and Jankowski, H. (2022).
\newblock Estimating the basic reproduction number at the beginning of an
  outbreak.
\newblock {\em PLoS ONE}, 17(6):e0269306.

\bibitem[Borremans et~al., 2016]{borremans2016}
Borremans, B., Hens, N., Beutels, P., Leirs, H., and Reijniers, J. (2016).
\newblock Estimating time of infection using prior serological and individual
  information can greatly improve incidence estimation of human and wildlife
  infections.
\newblock {\em PLoS computational biology}, 12(5):e1004882.

\bibitem[Cao et~al., 2021]{cao2021}
Cao, L., Zhao, S., Li, Q., Ling, L., Wu, W.~K., Zhang, L., Lou, J., Chong,
  M.~K., Chen, Z., Wong, E.~L., Zee, B. C.~Y., Chan, M. T.~V., Chan, P. K.~S.,
  and Wang, M.~H. (2021).
\newblock A {Bayesian} method for synthesizing multiple diagnostic outcomes of
  {COVID}-19 tests.
\newblock {\em Royal Society Open Science}, 8(9):201867.

\bibitem[Caprarelli and Fletcher, 2014]{caprarelli2014}
Caprarelli, G. and Fletcher, S. (2014).
\newblock A brief review of spatial analysis concepts and tools used for
  mapping, containment and risk modelling of infectious diseases and other
  illnesses.
\newblock {\em Parasitology}, 141(5):581--601.

\bibitem[Chandler et~al., 2021]{chandler2021}
Chandler, J.~C., Bevins, S.~N., Ellis, J.~W., Linder, T.~J., Tell, R.~M.,
  Jenkins-Moore, M., Root, J.~J., Lenoch, J.~B., Robbe-Austerman, S.,
  DeLiberto, T.~J., Gidlewski, T., Kim, T.~M., and Shriner, S.~A. (2021).
\newblock {SARS-CoV-2} exposure in wild white-tailed deer (\textit{Odocoileus
  virginianus}).
\newblock {\em Proceedings of the National Academy of Sciences},
  118(47):e2114828118.

\bibitem[Davis et~al., 2019]{davis2019}
Davis, A.~J., Kirby, J.~D., Chipman, R.~B., Nelson, K.~M., Xifara, T., Webb,
  C.~T., Wallace, R., Gilbert, A.~T., and Pepin, K.~M. (2019).
\newblock Not all surveillance data are created equal-—{A} multi-method
  dynamic occupancy approach to determine rabies elimination from wildlife.
\newblock {\em Journal of Applied Ecology}, 56(11):2551--2561.

\bibitem[{de Valpine} et~al., 2023]{nimble-software}
{de Valpine}, P., Paciorek, C., Turek, D., Michaud, N., Anderson-Bergman, C.,
  Obermeyer, F., Wehrhahn~Cortes, C., Rodr{\'i}guez, A., {Temple Lang}, D.,
  Zhang, W., Paganin, S., and Hug, J. (2023).
\newblock {NIMBLE}: {MCMC}, particle filtering, and programmable hierarchical
  modeling.

\bibitem[Dendukuri and Joseph, 2001]{dendukuri2001}
Dendukuri, N. and Joseph, L. (2001).
\newblock Bayesian approaches to modeling the conditional dependence between
  multiple diagnostic tests.
\newblock {\em Biometrics}, 57:158--167.

\bibitem[Faust et~al., 2018]{faust2018}
Faust, C.~L., McCallum, H.~I., Bloomfield, L. S.~P., Gottdenker, N.~L.,
  Gillespie, T.~R., Torney, C.~J., Dobson, A.~P., and Plowright, R.~K. (2018).
\newblock Pathogen spillover during land conversion.
\newblock {\em Ecology letters}, 21:471--483.

\bibitem[Featherstone et~al., 2022]{featherstone2022}
Featherstone, L.~A., Zhang, J.~M., Vaughan, T.~G., and Duchene, S. (2022).
\newblock Epidemiological inference from pathogen genomes: {A} review of
  phylodynamic models and applications.
\newblock {\em Virus Evolution}, 8(1):veac045.

\bibitem[Fox et~al., 2022]{fox2022}
Fox, S.~J., Lachmann, M., Tec, M., Pasco, R., Woody, S., Du, Z., Wang, X.,
  Ingle, T.~A., Javan, E., Dahan, M., Gaither, K., Escott, M.~E., Adler, S.~I.,
  Johnston, S.~C., Scott, J.~G., and Meyers, L.~A. (2022).
\newblock Real-time pandemic surveillance using hospital admissions and
  mobility data.
\newblock {\em Proceedings of the National Academy of Sciences},
  119(7):e2111870119.

\bibitem[Gelman et~al., 2020]{gelman2020}
Gelman, A., Carlin, J.~B., Dunson, D.~B., Vehtari, A., and Rubin, D.~B. (2020).
\newblock {\em Bayesian data analysis}.
\newblock Chapman and Hall/CRC, Boca Raton, FL, third ed. edition.

\bibitem[Gilbert et~al., 2013]{gilbert2013}
Gilbert, A.~T., Fooks, A., Hayman, D., Horton, D., M{\"u}ller, T., Plowright,
  R., Peel, A., Bowen, R., Wood, J., Mills, J., Cunningham, A., and CE, R.
  (2013).
\newblock Deciphering serology to understand the ecology of infectious diseases
  in wildlife.
\newblock {\em EcoHealth}, 10:298--313.

\bibitem[Gneiting and Raftery, 2007]{gneiting2007}
Gneiting, T. and Raftery, A.~E. (2007).
\newblock Strictly proper scoring rules, prediction, and estimation.
\newblock {\em Journal of the American Statistical Association},
  102(477):359--378.

\bibitem[Griffin, 2022]{griffin2022}
Griffin, D.~E. (2022).
\newblock Why does viral {RNA} sometimes persist after recovery from acute
  infections?
\newblock {\em PLoS biology}, 20(6):e3001687.

\bibitem[Habibzadeh et~al., 2022]{habibzadeh2022}
Habibzadeh, F., Habibzadeh, P., and Yadollahie, M. (2022).
\newblock The apparent prevalence, the true prevalence.
\newblock {\em Biochemia Medica}, 32(2):020101.

\bibitem[Hamer et~al., 2022]{hamer2022}
Hamer, S.~A., Nunez, C., Roundy, C.~M., Tang, W., Thomas, L., Richison, J.,
  Benn, J.~S., Auckland, L.~D., Hensley, T., Cook, W.~E., Pauvolid-Corr\^{e}a,
  A., and Hamer, G.~L. (2022).
\newblock Persistence of {SARS-CoV-2} neutralizing antibodies longer than 13
  months in naturally infected, captive white-tailed deer (\textit{Odocoileus
  virginianus}), {Texas}.
\newblock {\em Emerging microbes \& infections}, 11(1):2112--2115.

\bibitem[Helman et~al., 2020]{helman2020}
Helman, S.~K., Mummah, R.~O., Gostic, K.~M., Buhnerkempe, M.~G., Prager, K.~C.,
  and Lloyd-Smith, J.~O. (2020).
\newblock Estimating prevalence and test accuracy in disease ecology: How
  {Bayesian} latent class analysis can boost or bias imperfect test results.
\newblock {\em Ecology and Evolution}, 10(14):7221--7232.

\bibitem[Hewitt et~al., 2023]{hewitt2023}
Hewitt, J., Wilson-Henjum, G., Collins, D.~T., Linder, T.~J., Lenoch, J.~B.,
  Heale, J.~D., Quintanal, C.~A., Pleszewski, R., McBride, D.~S., Bowman,
  A.~S., Chandler, J.~C., Shriner, S.~A., Bevins, S.~N., Kohler, D.~J.,
  Chipman, R.~B., Gosser, A.~L., Bergman, D.~L., DeLiberto, T.~J., and Pepin,
  K.~M. (2023).
\newblock Epidemiological dynamics of {SARS-CoV-2} in white-tailed deer.
\newblock {\em (Submitted)}.

\bibitem[Irvine et~al., 2007]{irvine2007}
Irvine, K.~M., Gitelman, A.~I., and Hoeting, J.~A. (2007).
\newblock Spatial designs and properties of spatial correlation: effects on
  covariance estimation.
\newblock {\em Journal of Agricultural, Biological, and Environmental
  Statistics}, 12:450--469.

\bibitem[Jia et~al., 2020]{jia2020}
Jia, B., Colling, A., Stallknecht, D.~E., Blehert, D., Bingham, J., Crossley,
  B., Eagles, D., and Gardner, I.~A. (2020).
\newblock Validation of laboratory tests for infectious diseases in wild
  mammals: review and recommendations.
\newblock {\em Journal of Veterinary Diagnostic Investigation}, 32(6):776--792.

\bibitem[Joynt and Wu, 2020]{joynt2020}
Joynt, G.~M. and Wu, W.~K. (2020).
\newblock Understanding {COVID}-19: what does viral {RNA} load really mean?
\newblock {\em The Lancet Infectious Diseases}, 20(6):635--636.

\bibitem[King et~al., 2022]{king2022}
King, A.~A., Lin, Q., and Ionides, E.~L. (2022).
\newblock Markov genealogy processes.
\newblock {\em Theoretical population biology}, 143:77--91.

\bibitem[Lau et~al., 2015]{lau2015}
Lau, M.~S., Marion, G., Streftaris, G., and Gibson, G. (2015).
\newblock A systematic {Bayesian} integration of epidemiological and genetic
  data.
\newblock {\em PLoS computational biology}, 11(11):e1004633.

\bibitem[Lo~Iacono et~al., 2016]{lo2016}
Lo~Iacono, G., Cunningham, A.~A., Fichet-Calvet, E., Garry, R.~F., Grant,
  D.~S., Leach, M., Moses, L.~M., Nichols, G., Schieffelin, J.~S., Shaffer,
  J.~G., Webb, C.~T., and Wood, J. L.~N. (2016).
\newblock A unified framework for the infection dynamics of zoonotic spillover
  and spread.
\newblock {\em PLoS neglected tropical diseases}, 10(9):e0004957.

\bibitem[Manlove et~al., 2022]{manlove2022}
Manlove, K., Wilber, M., White, L., Bastille-Rousseau, G., Yang, A.,
  Gilbertson, M., Craft, M., Cross, P., Wittemyer, G., and Pepin, K. (2022).
\newblock Defining an epidemiological landscape by connecting host movement to
  pathogen transmission.
\newblock {\em Ecology Letters}, 25:1760--1782.

\bibitem[Martcheva, 2015]{martcheva2015}
Martcheva, M. (2015).
\newblock {\em An introduction to mathematical epidemiology}.
\newblock Springer, New York, NY.

\bibitem[Martins et~al., 2022]{martins2022}
Martins, M., Boggiatto, P.~M., Buckley, A., Cassmann, E.~D., Falkenberg, S.,
  Caserta, L.~C., Fernandes, M. H.~V., Kanipe, C., Lager, K., Palmer, M.~V.,
  and Diel, D.~G. (2022).
\newblock From deer-to-deer: {SARS-CoV-2} is efficiently transmitted and
  presents broad tissue tropism and replication sites in white-tailed deer.
\newblock {\em PLoS pathogens}, 18(3):e1010197.

\bibitem[McBride et~al., 2023]{mcbride2023}
McBride, D.~S., Garushyants, S.~K., Franks, J., Magee, A.~F., Overend, S.~H.,
  Huey, D., Williams, A.~M., Faith, S.~A., Kandeil, A., Trifkovic, S., Miller,
  L., Jeevan, T., Patel, A., Nolting, J.~M., Tonkovich, M.~J., Genders, J.~T.,
  Montoney, A.~J., Kasnyik, K., Linder, T.~J., Bevins, S.~N., Lenoch, J.~B.,
  Chandler, J.~C., DeLiberto, T.~J., Koonin, E.~V., Suchard, M.~A., Lemey, P.,
  Webby, R.~J., Nelson, M.~I., and Bowman, A.~S. (2023).
\newblock Accelerated evolution of {SARS}-{CoV}-2 in free-ranging white-tailed
  deer.
\newblock {\em Nature Communications}, 14:5105.

\bibitem[Meentemeyer et~al., 2012]{meentemeyer2012}
Meentemeyer, R.~K., Haas, S.~E., and V{\'a}clav{\'\i}k, T. (2012).
\newblock Landscape epidemiology of emerging infectious diseases in natural and
  human-altered ecosystems.
\newblock {\em Annual review of Phytopathology}, 50:379--402.

\bibitem[Mysterud et~al., 2023]{mysterud2023}
Mysterud, A., Viljugrein, H., Hopp, P., Andersen, R., Bakka, H., Benestad,
  S.~L., Madslien, K., Moldal, T., Rauset, G.~R., Strand, O., Tran, L.,
  Vik{\o}ren, T., and V{\aa}ge, J. (2023).
\newblock Challenges and opportunities using hunters to monitor chronic wasting
  disease among wild reindeer in the digital era.
\newblock {\em Ecological Solutions and Evidence}, 4:e12203.

\bibitem[Nichols et~al., 2021]{nichols2021}
Nichols, J.~D., Bogich, T.~L., Howerton, E., Bj{\o}rnstad, O.~N., Borchering,
  R.~K., Ferrari, M., Haran, M., Jewell, C., Pepin, K.~M., Probert, W. J.~M.,
  Pulliam, J. R.~C., Runge, M.~C., Tildesley, M., Viboud, C., and Shea, K.
  (2021).
\newblock Strategic testing approaches for targeted disease monitoring can be
  used to inform pandemic decision-making.
\newblock {\em PLoS biology}, 19(6):e3001307.

\bibitem[Nielsen et~al., 2007]{nielsen2007}
Nielsen, L.~R., van~den Borne, B., and van Schaik, G. (2007).
\newblock Salmonella {Dublin} infection in young dairy calves: transmission
  parameters estimated from field data and an {SIR}-model.
\newblock {\em Preventive Veterinary Medicine}, 79:46--58.

\bibitem[O'Dea et~al., 2014]{odea2014}
O'Dea, E.~B., Pepin, K.~M., Lopman, B.~A., and Wilke, C.~O. (2014).
\newblock Fitting outbreak models to data from many small norovirus outbreaks.
\newblock {\em Epidemics}, 6:18--29.

\bibitem[Palmer et~al., 2021]{palmer2021}
Palmer, M.~V., Martins, M., Falkenberg, S., Buckley, A., Caserta, L.~C.,
  Mitchell, P.~K., Cassmann, E.~D., Rollins, A., Zylich, N.~C., Renshaw, R.~W.,
  Guarino, C., Wagner, B., Lager, K., and Diel, D.~G. (2021).
\newblock Susceptibility of white-tailed deer (\textit{Odocoileus virginianus})
  to {SARS-CoV-2}.
\newblock {\em Journal of Virology}, 95(11):e00083--21.

\bibitem[Pepin et~al., 2017]{pepin2017}
Pepin, K.~M., Kay, S.~L., Golas, B.~D., Shriner, S.~S., Gilbert, A.~T., Miller,
  R.~S., Graham, A.~L., Riley, S., Cross, P.~C., Samuel, M.~D., et~al. (2017).
\newblock Inferring infection hazard in wildlife populations by linking data
  across individual and population scales.
\newblock {\em Ecology letters}, 20(3):275--292.

\bibitem[Pepin et~al., 2019]{pepin2019}
Pepin, K.~M., Pedersen, K., Wan, X.-F., Cunningham, F.~L., Webb, C.~T., and
  Wilber, M.~Q. (2019).
\newblock Individual-level antibody dynamics reveal potential drivers of
  influenza {A} seasonality in wild pig populations.
\newblock {\em Integrative and Comparative Biology}, 59(5):1231--1242.

\bibitem[Perera et~al., 2021]{perera2021}
Perera, R.~A., Ko, R., Tsang, O.~T., Hui, D.~S., Kwan, M.~Y., Brackman, C.~J.,
  To, E.~M., Yen, H.-l., Leung, K., Cheng, S.~M., Chan, K.~H., Chan, K. C.~K.,
  Li, K.-C., Saif, L., Barrs, Vanessa R.~Wu, J.~T., Sit, T. H.~C., Poon, L.
  L.~M., and Peiris, M. (2021).
\newblock Evaluation of a {SARS-CoV-2} surrogate virus neutralization test for
  detection of antibody in human, canine, cat, and hamster sera.
\newblock {\em Journal of Clinical Microbiology}, 59(2):e02504--20.

\bibitem[Podg{\'o}rski et~al., 2020]{podgorski2020}
Podg{\'o}rski, T., Borowik, T., {\L}yjak, M., and Wo{\'z}niakowski, G. (2020).
\newblock Spatial epidemiology of {African} swine fever: {Host}, landscape and
  anthropogenic drivers of disease occurrence in wild boar.
\newblock {\em Preventive Veterinary Medicine}, 177:104691.

\bibitem[Prager et~al., 2020]{prager2020}
Prager, K., Buhnerkempe, M.~G., Greig, D.~J., Orr, A.~J., Jensen, E.~D., Gomez,
  F., Galloway, R.~L., Wu, Q., Gulland, F.~M., and Lloyd-Smith, J.~O. (2020).
\newblock Linking longitudinal and cross-sectional biomarker data to understand
  host-pathogen dynamics: {Leptospira} in {California} sea lions
  (\emph{Zalophus californianus}) as a case study.
\newblock {\em PLOS Neglected Tropical Diseases}, 14(6):e0008407.

\bibitem[Ramakrishnan et~al., 2021]{ramakrishnan2021}
Ramakrishnan, R.~K., Kashour, T., Hamid, Q., Halwani, R., and Tleyjeh, I.~M.
  (2021).
\newblock Unraveling the mystery surrounding post-acute sequelae of {COVID}-19.
\newblock {\em Frontiers in Immunology}, 12:686029.

\bibitem[Rogan and Gladen, 1978]{rogan1978}
Rogan, W.~J. and Gladen, B. (1978).
\newblock Estimating prevalence from the results of a screening test.
\newblock {\em American Journal of Epidemiology}, 107(1):71--76.

\bibitem[Simonsen et~al., 2009]{simonsen2009}
Simonsen, J., M{\o}lbak, K., Falkenhorst, G., Krogfelt, K., Linneberg, A., and
  Teunis, P. (2009).
\newblock Estimation of incidences of infectious diseases based on antibody
  measurements.
\newblock {\em Statistics in Medicine}, 28(14):1882--1895.

\bibitem[Smith et~al., 2017]{smith2017}
Smith, R.~A., Ionides, E.~L., and King, A.~A. (2017).
\newblock Infectious disease dynamics inferred from genetic data via sequential
  {Monte} {Carlo}.
\newblock {\em Molecular Bbiology and Evolution}, 34(8):2065--2084.

\bibitem[Subramanian et~al., 2021]{subramanian2021}
Subramanian, R., He, Q., and Pascual, M. (2021).
\newblock Quantifying asymptomatic infection and transmission of {COVID}-19 in
  {New} {York} {City} using observed cases, serology, and testing capacity.
\newblock {\em Proceedings of the National Academy of Sciences},
  118(9):e2019716118.

\bibitem[Tabak et~al., 2019]{tabak2019}
Tabak, M.~A., Pedersen, K., and Miller, R.~S. (2019).
\newblock Detection error influences both temporal seroprevalence predictions
  and risk factors associations in wildlife disease models.
\newblock {\em Ecology and evolution}, 9(18):10404--10414.

\bibitem[Tan et~al., 2020]{tan2020}
Tan, C.~W., Chia, W.~N., Qin, X., Liu, P., Chen, M. I.-C., Tiu, C., Hu, Z.,
  Chen, V. C.-W., Young, B.~E., Sia, W.~R., Tan, Y.-J., Foo, R., Yi, Y., Lye,
  D.~C., Anderson, D.~E., and Wang, L.-F. (2020).
\newblock A {SARS-CoV-2} surrogate virus neutralization test based on
  antibody-mediated blockage of {ACE2}--spike protein--protein interaction.
\newblock {\em Nature biotechnology}, 38(9):1073--1078.

\bibitem[Teunis et~al., 2012]{teunis2012}
Teunis, P., Van~Eijkeren, J., Ang, C., van Duynhoven, Y., Simonsen, J., Strid,
  M., and van Pelt, W. (2012).
\newblock Biomarker dynamics: estimating infection rates from serological data.
\newblock {\em Statistics in medicine}, 31(20):2240--2248.

\bibitem[Vinh and Boni, 2015]{vinh2015}
Vinh, D.~N. and Boni, M.~F. (2015).
\newblock Statistical identifiability and sample size calculations for serial
  seroepidemiology.
\newblock {\em Epidemics}, 12:30--39.

\bibitem[Volz et~al., 2009]{volz2009}
Volz, E.~M., Kosakovsky~Pond, S.~L., Ward, M.~J., Leigh~Brown, A.~J., and
  Frost, S.~D. (2009).
\newblock Phylodynamics of infectious disease epidemics.
\newblock {\em Genetics}, 183(4):1421--1430.

\bibitem[Volz and Siveroni, 2018]{volz2018}
Volz, E.~M. and Siveroni, I. (2018).
\newblock Bayesian phylodynamic inference with complex models.
\newblock {\em PLoS computational biology}, 14(11):e1006546.

\bibitem[Wilber et~al., 2020]{wilber2020}
Wilber, M.~Q., Webb, C.~T., Cunningham, F.~L., Pedersen, K., Wan, X.-F., and
  Pepin, K.~M. (2020).
\newblock Inferring seasonal infection risk at population and regional scales
  from serology samples.
\newblock {\em Ecology}, 101(1):e02882.

\end{thebibliography}

\end{document}